\documentclass[journal=jacsat,manuscript=article]{achemso}

\usepackage[version=3]{mhchem} 
\usepackage[T1]{fontenc}       
\usepackage[squaren]{SIunits}
\usepackage{mciteplus}

\author{Olivier Maillet}
\affiliation{Universit\'e Grenoble Alpes, CNRS Institut N\'eel, BP 166, 38042 Grenoble Cedex 9, France}
\author{Xin Zhou}
\affiliation{Universit\'e Grenoble Alpes, CNRS Institut N\'eel, BP 166, 38042 Grenoble Cedex 9, France\\
Now at: Institut d'Electronique, de Micro\'electronique et de Nanotechnologie (IEMN), Univ. Lille \& CNRS, 59000 Lille, France}
\author{Rasul R. Gazizulin}
\affiliation{Universit\'e Grenoble Alpes, CNRS Institut N\'eel, BP 166, 38042 Grenoble Cedex 9, France}
\author{Rob Ilic} 
\affiliation{Center for Nanoscale Science and Technology, National Institute of Standards and Technology, Gaithersburg, Maryland 20899, USA}
\author{Jeevak M. Parpia}
\affiliation{Department of Physics, Cornell University, Ithaca, New York 14853, USA}
\author{Olivier Bourgeois}
\affiliation{Universit\'e Grenoble Alpes, CNRS Institut N\'eel, BP 166, 38042 Grenoble Cedex 9, France}
\author{Andrew D. Fefferman}
\affiliation{Universit\'e Grenoble Alpes, CNRS Institut N\'eel, BP 166, 38042 Grenoble Cedex 9, France}
\author{Eddy Collin}
\affiliation{Universit\'e Grenoble Alpes, CNRS Institut N\'eel, BP 166, 38042 Grenoble Cedex 9, France}
\email{eddy.collin@neel.cnrs.fr}
\phone{+33 4 76 88 78 31}
\title[An \textsf{achemso} demo]
  {Measuring Frequency Fluctuations in Nonlinear Nanomechanical Resonators}

\abbreviations{IR,NMR,UV}

\begin{document}

\begin{tocentry}

Some journals require a graphical entry for the Table of Contents.
This should be laid out ``print ready'' so that the sizing of the
text is correct.

Inside the \texttt{tocentry} environment, the font used is Helvetica
8\,pt, as required by \emph{Journal of the American Chemical
Society}.

The surrounding frame is 9\,cm by 3.5\,cm, which is the maximum
permitted for  \emph{Journal of the American Chemical Society}
graphical table of content entries. The box will not resize if the
content is too big: instead it will overflow the edge of the box.

This box and the associated title will always be printed on a
separate page at the end of the document.

\end{tocentry}

\begin{abstract}

Advances in nanomechanics within recent years have demonstrated an 
always expanding range of devices, from top-down structures to appealing bottom-up MoS$_2$ and graphene membranes, used for both sensing and component-oriented applications.
One of the main concerns in all of these devices is frequency noise, which ultimately limits their applicability. This issue has attracted a lot of attention recently, and the origin of this noise remains elusive up to date. 
In this Letter we present a very simple technique to measure frequency noise in nonlinear mechanical devices, based on the presence of bistability. It is illustrated on silicon-nitride high-stress doubly-clamped beams, in a cryogenic environment.
We report on the same $T/f$ dependence of the frequency noise power spectra as reported in the literature. 
But we also find unexpected {\it damping fluctuations}, amplified in the vicinity of the bifurcation points; this effect is clearly distinct from already reported nonlinear dephasing, 
 and poses a fundamental limit on the measurement of bifurcation frequencies.
The technique is further applied to the measurement of frequency noise as a function of mode number, within the same device. 
The relative frequency noise for the fundamental flexure $\delta f/f_0$ lies in the range $0.5 - 0.01~$ppm (consistent with literature for cryogenic MHz devices
), and decreases with mode number in the range studied. 
The technique can be applied to {\it any types} of nano-mechanical structures, enabling progresses towards the understanding of intrinsic sources of noise in these devices. 


\end{abstract}



Within the past decade Nano-Electro-Mechanical-Systems (NEMS) have developed with a broad range of applications extending from physics to engineering.
In the first place, their size makes them very sensitive transducers of force \cite{moser,liroukes}.
This had been demonstrated {\it e.g.} in the seminal work of D. Rugar {\it et al.} in which a cantilever loaded by a magnetic tip reached a detection sensitivity corresponding to the force exerted by a single electronic spin at a distance of about 100$~$nm \cite{rugar}.
More recently, NEMS have been applied to the detection of small quantities of matter (mass spectroscopy), with precision reaching the single proton \cite{chaste}. Nowadays, even the quantum nature of the mechanical degree of freedom is exploited for quantum information processing \cite{quantum}.

In all applications, the quality of the device is intrinsically linked to its level of displayed noise \cite{clelandnoise}. Specifically, 
frequency noise in NEMS appears to be a key limiting parameter whose physical origin is still unknown \cite{hentz}.
Besides, only few quantitative experimental studies are available in the literature \cite{hentz,bachtolddykman,greywall,PLL}, especially at low temperatures \cite{tang}. The nonlinear frequency noise reported for carbon-based systems \cite{venstra} is one of the most striking results, revealing the complex nature that the underlying mechanisms can possess. 

Frequency noise (or phase noise \cite{clelandnoise,greywall,tang}) can be understood in terms of {\it pure dephasing}\cite{venstra}, making an analogy with Nuclear Magnetic Resonance (NMR); and its impact on a mechanical resonance can be modeled experimentally by means of engineered frequency fluctuations \cite{NJPus}. 
The physical origin of intrinsic frequency noise is indeed still elusive, since all identified mechanisms studied explicitly display much weaker contributions than the reported experimental values: adsorption-desorption/mobility of surface atoms \cite{atalayadykman}, experimentally modeled under a Xe flow \cite{diffuseRoukes}, or the nonlinear transduction of Brownian motion \cite{zhanganddyk,PRBBrownian}. 
These efforts in understanding the microscopic mechanisms at work in mechanical dephasing are accompanied by theoretical support. The nonlinear dephasing/damping has been proposed to originate in nonlinear phononic interactions between the low frequency mechanical modes and thermal phonons \cite{nonlindephase}. 
Finally, a common speculation reported in the literature is that frequency noise is related to defects \cite{clelandnoise,tang,dutta}, which can be either extrinsic or constitutive of the material (like in a glass). 
The presence of these so-called Two-Level Systems (TLS) is also proposed to explain
damping mechanisms in NEMS \cite{TLSnems1,TLSnems2}, and have been shown recently to 
lead to peculiar features (especially in the noise) for mesoscopic systems such as quantum bits and NEMS \cite{tlspaper}.
 
Properly measuring frequency noise is not easy; a neat technique presented in the literature relies on cross-correlations present in the two signals of a dual-tone scheme \cite{hentz}. Moreover, in the spectral domain dephasing and damping are mixed \cite{graphene,venstra,NJPus}. 
In order to separate the contributions, one has to use {\it both} spectral-domain and time-domain measurements \cite{venstra,NJPus}. All of these techniques may not be well suited for large amplitude signals (especially when the system becomes bistable), preventing the exploration of the nonlinear range where nonlinear damping/dephasing may dominate. 

In this Letter we present a method based on bifurcation enabling a very simple measurement of frequency noise in nonlinear bistable resonators. Building on this method, we characterize the intrinsic frequency noise of high-stress Silicon Nitride (SiN) doubly-clamped beams in cryogenic environment (form 1.4$~$K to 30$~$K). 
In particular, we study the three first symmetric modes ($n=1,3,5$) of one of our devices, and demonstrate the compatibility of our results with existing literature. The temperature-dependence is indeed similar to Ref. \cite{tang}, but we find an unexpected {\it damping noise} which is amplified through the bifurcation measurement.
This result is distinct from the reported nonlinear phase noise of Ref. \cite{venstra} in which the device was {\it not} bistable. 
The phenomenon seems to be generic, and we discuss it in the framework of the 
TLS model. 
Note that our results demonstrate the existence of an ultimate limit to the frequency resolution of bifurcation points in nonlinear mechanical systems.
					
		\begin{figure}[h!]
	\includegraphics[width=16cm]{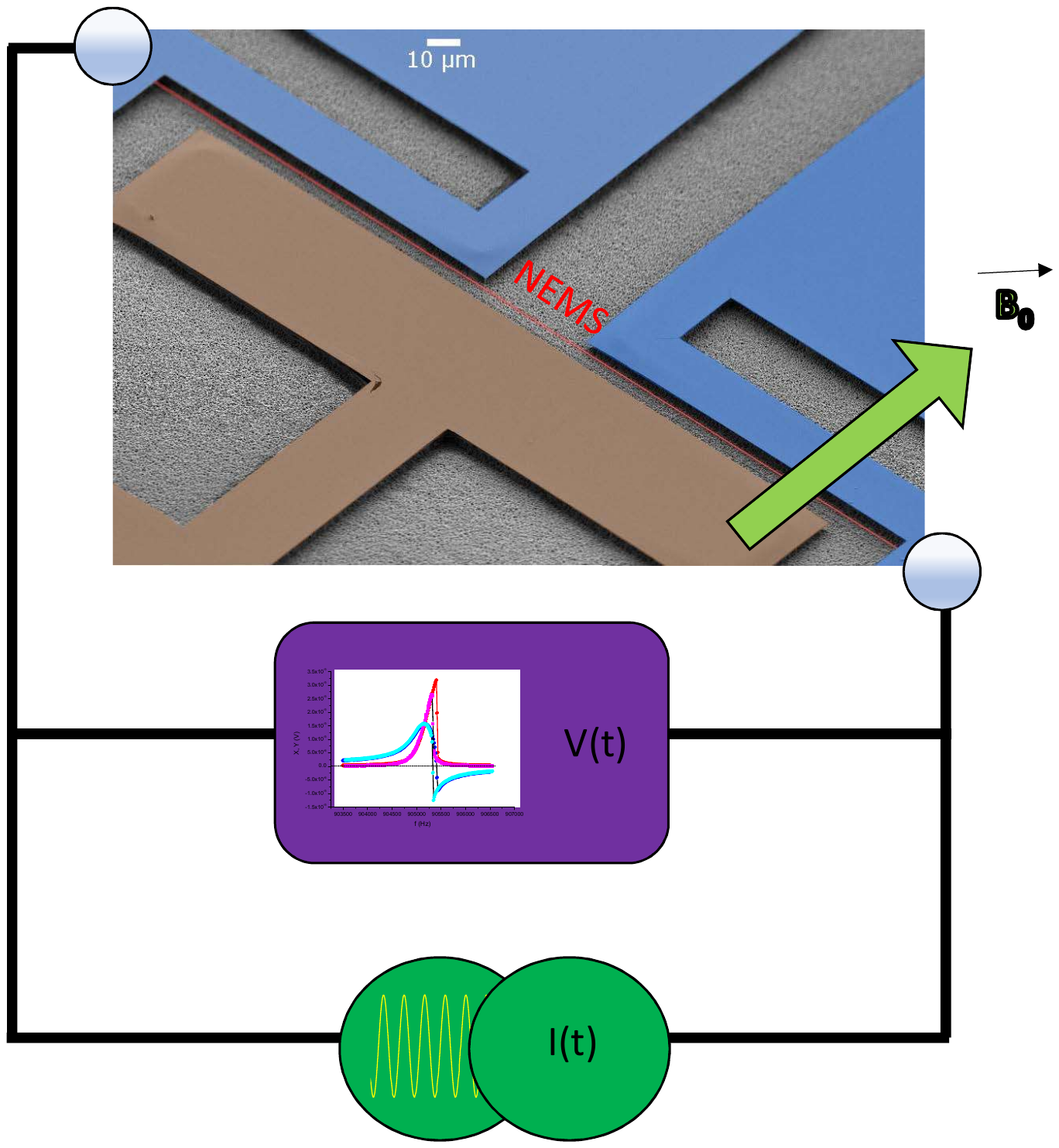} 			
			\caption{\small{Device and setup. 
			 SEM image of the 250$~\mu$m device measured in this work. The gate electrode (brown) is not used here. The actual NEMS device is the red-colored string in between the two (light blue) electrodes.
			The lock-in detector (violet), magnetic field and drive generator (in green) are also depicted in a schematic fashion to illustrate the magnetomotive technique.
			}}
			\label{fig_setup}
		\end{figure}

\section{Results and discussion}

A typical doubly-clamped NEMS device used in our work is shown in Fig. \ref{fig_setup}.
It consists of a 100$~$nm thick SiN device covered with 30$~$nm of Al. The width of the beam is 300$~$nm and the length $L=250~\mu$m. 
Another similar sample of $L=15~\mu$m has been characterized. The beams store about 1$~$GPa of tensile stress, and we define $A$ to be their rectangular cross-section. For fabrication details see Methods below.
The device is placed in a $^4$He cryostat with temperature $T_0$ regulated between 1.4$~$K and 30$~$K, under cryogenic vacuum $\leq 10^{-6}~$mbar. The motion of the beam is driven and detected by means of the magnetomotive scheme \cite{clelandsensors,RSI}. For experimental details see Methods below. 
			
		\begin{figure}[h!]
	 			\includegraphics[width=8.1 cm]{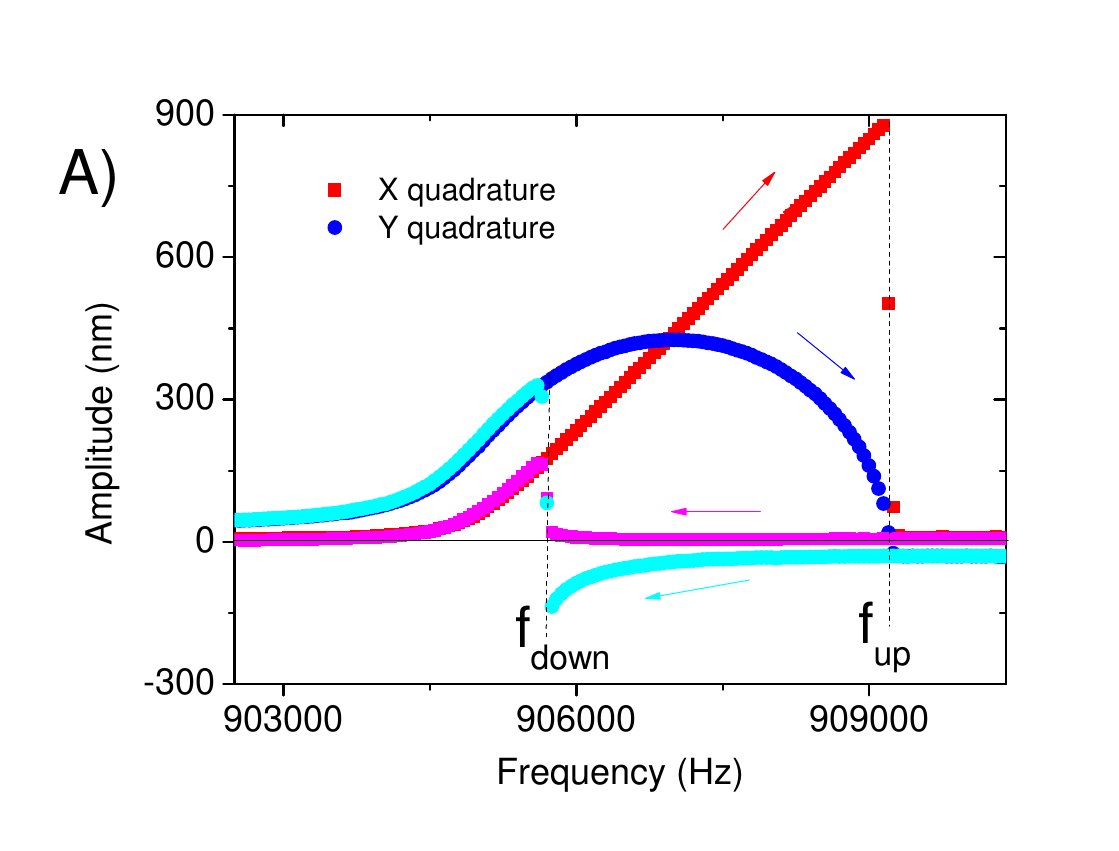}  \includegraphics[width=8.1 cm]{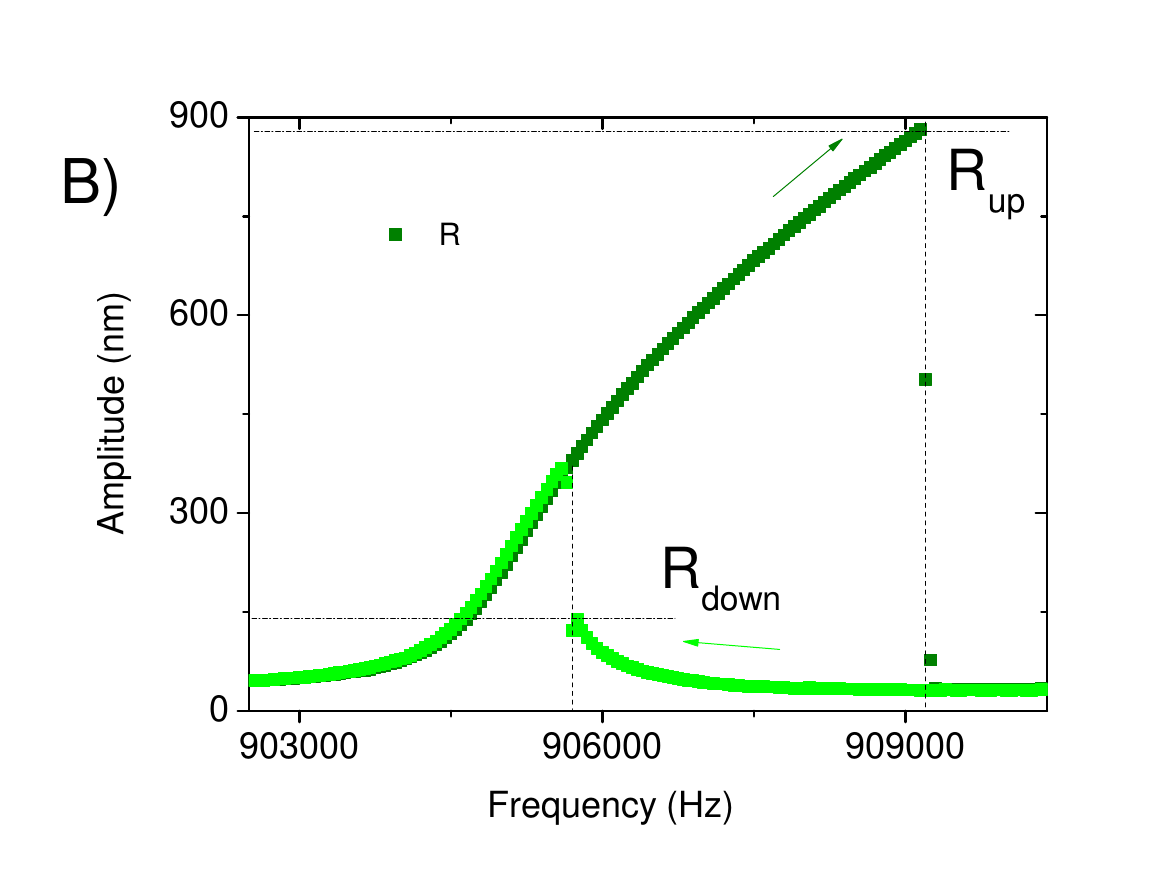}
			\caption{\small{Nonlinear (Duffing) resonance. 
			 A) Duffing resonance line (X and Y quadratures) measured on the 250$~\mu$m device at 4.2$~$K in vacuum, for a drive force of 81$~$pN, in a 1$~$T field. The directions of frequency sweeps are depicted by arrows. Vertical dashed lines indicate the two bifurcation points $f_{up,down}$.
			 B) Amplitude $R=\sqrt{X^2+Y^2}$ corresponding to A). Bifurcation points are indicated with their amplitudes $R_{up,down}$. 
			}}
			\label{fig_lines}
		\end{figure}

A Laplace force $F(t)=F_0 \cos(2 \pi f\, t)$ with $F_0 \propto I_0 L B_0$ is created with a static in-plane magnetic field $B_0$ and an AC current $I_0$ fed into the metallic layer (Fig. \ref{fig_setup}). Fields $B_0$ of the order of 1$~$T, and currents $I_0$ up to 0.5$~\mu$A have been used.
The detected signal is the induced voltage $V(t)$ proportional to velocity. It is measured with a lock-in from which we can obtain the two quadratures $X,Y$ of the motion. We call $R=\sqrt{X^2+Y^2}$ the amplitude of the motion (at a given frequency), defined in meters peak.
For all the $T_0$, $B_0$ settings used in the present work, the Al layer was not superconducting.
A key feature of the magnetomotive scheme is that it enables the ability to tune the $Q$ factor of the detected resonances \cite{clelandsensors}: this is the so-called {\it loading} effect.

At low drives, in the linear regime, the quality factor of the resonance $Q=f_0/\Delta f$ is defined from the linewidth $\Delta f$ and the resonance frequency $f_0$ of the mode under study. We consider here only high-$Q$ resonances $Q \gg 1$. In this limit, the $X$ peak is a simple Lorentzian, whose full-width-at-half-height gives $\Delta f$. 
For large excitation forces, our doubly-clamped beams' mechanical modes behave as almost ideal Duffing resonators \cite{lifshitz,roukes}. A typical Duffing resonance is shown in Fig. \ref{fig_lines}. 
The maximum of the resonance shifts with motion amplitude as $f_{max}=f_0 + \beta R_{max}^2$. $\beta$ is the so-called Duffing parameter. We assume $\beta >0$, but the case $\beta <0$ is straightforward to adapt.
$R_{max}$ is the maximum amplitude of motion; it always satisfies $R_{max} = F_0 Q/k_0$ with very good accuracy \cite{PRBus}. $k_0$ is the mode's spring constant with $f_0 =\frac{1}{2 \pi} \sqrt{k_0/m_0}$ and $m_0$ the mode mass.  
In the nonlinear Duffing regime, a damping parameter $\Delta f$ can still be defined from the $Q$ factor deduced from the peak height $R_{max}$. When frequency noise is negligible, the so-called decoherence time $T_2 = 1/(\pi \Delta f)$ defined from such frequency-domain measurements is just equal to $T_1$, the relaxation time of the amplitude $R$ in time-domain \cite{NJPus}.

		\begin{figure}[h!]
	 			\includegraphics[width=10. cm]{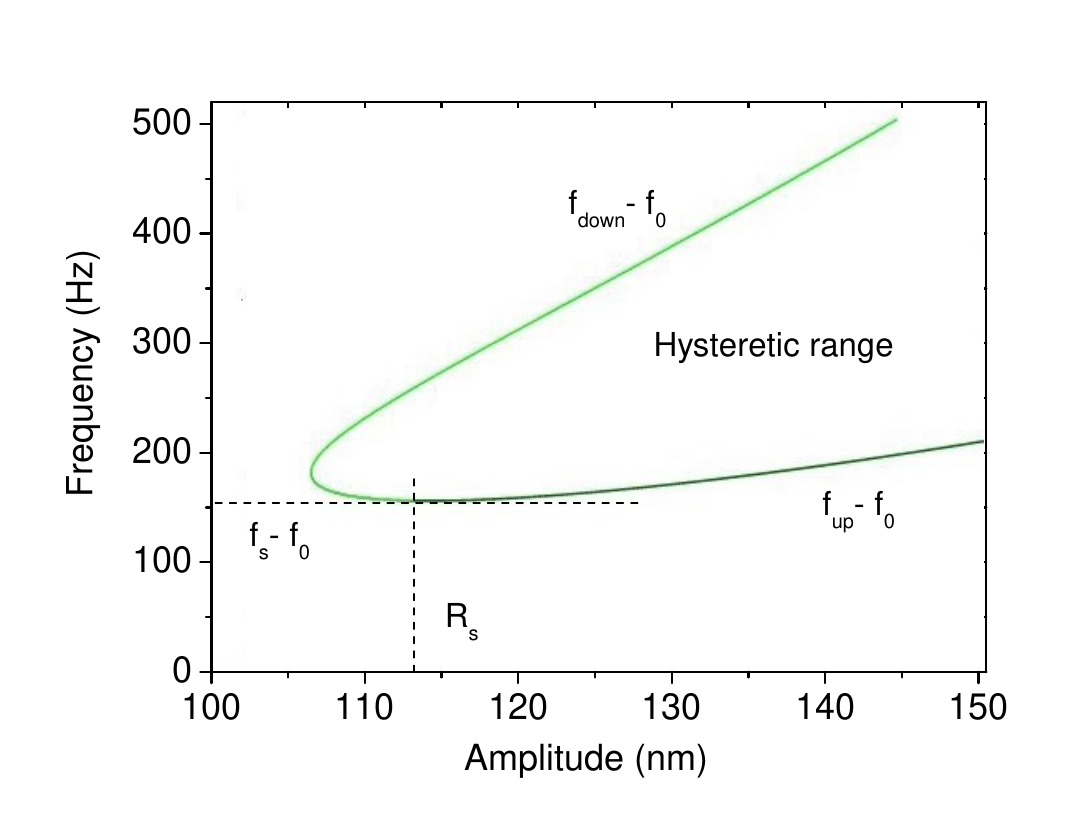}
			\caption{\small{Bifurcation branches. 
			Calculated bifurcation frequencies from Eqs. (\ref{fup}-\ref{fdown}) for the 250$~\mu$m device and magnetomotively-loaded $Q$ of $5\,000$ (1$~$T field).
			}}
			\label{fig_branches}
		\end{figure}
		
When $R_{max}\geq \frac{2}{\sqrt{3}} R_s$ a response hysteresis opens \cite{landau} (see Fig. \ref{fig_lines}). The point in the $(R,f)$ space at which this begins is called the {\it spinodal point}, with $R_s= \frac{1}{3^{1/4}} \sqrt{\frac{\Delta f}{\beta}}$. Beyond this point, two stable states coexist in a range of frequencies, with the system jumping from one to the other at $f_{up}$ (in an upward frequency sweep) and $f_{down}$ (for downward). These are the two {\it bifurcation points}. The maximum amplitude of the resonance is reached only on the upper branch.
These frequencies write explicitly:
\begin{eqnarray}
f_{up}   &=& f_0 + 2 \beta R_{up}^2 - \sqrt{\left( \beta R_{up}^2 \right)^2 - \frac{1}{4} \Delta f^2} \,\,\,\,\,\mbox{for}\,\,\,\,\, R_{up} \geq R_s, \label{fup} \\
&\mbox{and:} & \nonumber \\
f_{down} &=& f_0 + 2 \beta R_{down}^2 - \sqrt{\left( \beta R_{down}^2 \right)^2 - \frac{1}{4} \Delta f^2} \,\,\,\,\,\mbox{for}\,\,\,\,\, \frac{3^{1/4}}{\sqrt{2}} R_s < R_{down} \leq R_s  \,\,\mbox{ \it (Case 1)}, \nonumber  \\
f_{down} &=& f_0 + 2 \beta R_{down}^2 + \sqrt{\left( \beta R_{down}^2 \right)^2 - \frac{1}{4} \Delta f^2} \,\,\,\,\,\mbox{for}\,\,\,\,\, R_{down} \geq \frac{3^{1/4}}{\sqrt{2}} R_s  \,\,\mbox{ \it (Case 2)}. \label{fdown}
\end{eqnarray}
These functions are displayed in Fig. \ref{fig_branches}.
As we increase the driving force $F_0$, the maximum amplitude $R_{max}$ linearly increases while the peak position shifts quadratically. The upper bifurcation point $f_{up}$ then shifts monotonically towards higher frequencies, with a monotonically increasing amplitude $R_{up}$ when $F_0$ is increased. 
On the other hand, the lower bifurcation point $f_{down}$ has first an amplitude $R_{down}$ that decreases ($f_{down}$ being given by {\it Case 1} above), and then it increases again ($f_{down}$ being then defined through {\it Case 2}).
At the spinodal point $R_s$, $f_{up}=f_{down}=f_s = f_0 + \frac{\sqrt{3}}{2} \Delta f$.

	\begin{figure}[h!]
			 \includegraphics[width=8cm]{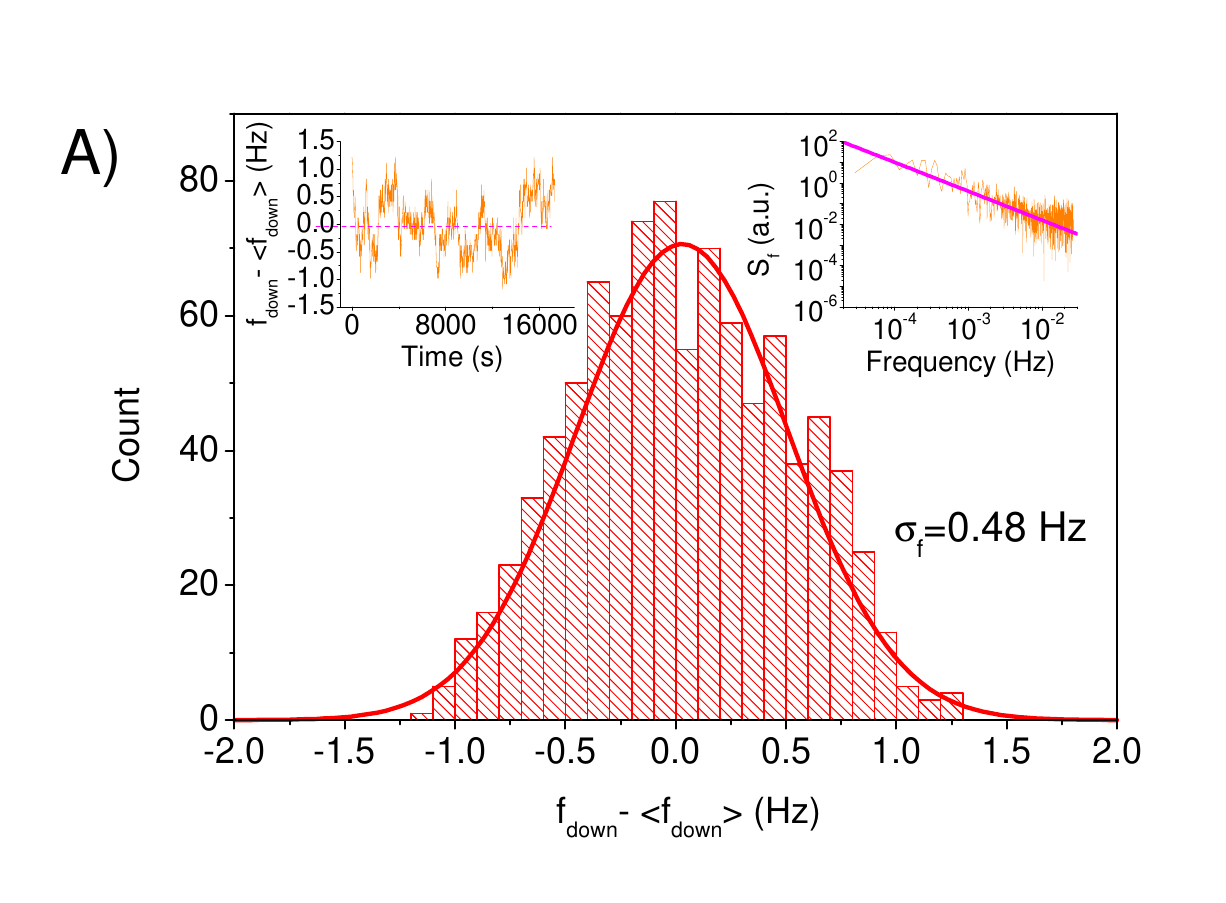}	 \includegraphics[width=8.37cm]{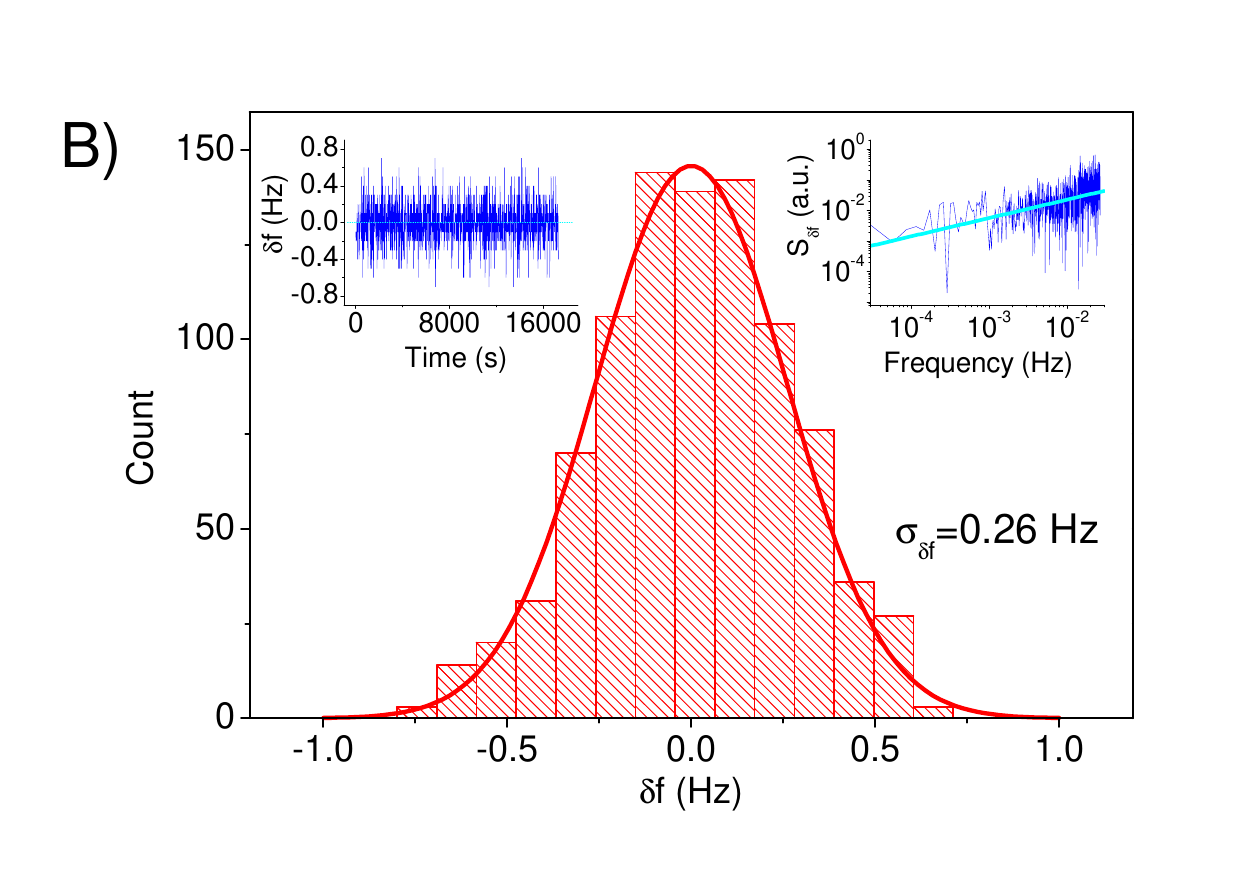}
			\caption{\small{Statistics on frequency at low amplitudes. 
			A): histogram obtained on the $f_{down}$ relaxation point of Fig. \ref{fig_lines} (left inset: actual frequency time-trace; right inset: power spectrum of the frequency fluctuations, displaying $1/f$-type structure). B): histogram obtained on the frequency jumps $\delta f$ computed from the $f_{down}$ time-trace (left inset: $\delta f$ time-trace; right inset: power spectrum). The lines are Gaussian fits, and the power-law dependencies of the spectra are discussed in the text.}}
			\label{fig_lowstats}
		\end{figure}	
	
The method we present builds on the work by Aldridge and Cleland \cite{AldridgeCleland}: the bifurcation positions are essentially {\it arbitrarily well defined} (in the sense that in an ideal system only the thermal activation of the bifurcation process will limit the stability), and can be used for sensing/amplifying. We thus devise a technique enabling the characterization of frequency fluctuations themselves; indeed, the imprint of frequency fluctuations had been reported earlier in noise-induced bifurcation relaxation \cite{martial}.
We show in Fig. \ref{fig_lowstats} histograms obtained on the $f_{down}$ frequency position of the resonance of Fig. \ref{fig_lines}.			
They are measured by ramping the frequency down from above $f_{up}$ at constant speed, and measuring the switching to the higher branch through a threshold detection. We repeat this protocol typically 1000$~$times to acquire enough statistics (see Fig. \ref{fig_lowstats} graph A left inset). The histogram obtained directly from the frequency time-trace is then fit to a Gaussian (of standard deviation $\sigma_{f}$, graph A in Fig. \ref{fig_lowstats}), while the power spectrum $S_{f}(f)$ of the fluctuations is also computed (defined as the FFT of the auto-correlation function). It displays a $1/f$-type structure (see Fig. \ref{fig_lowstats} A right inset).

Some precautions have to be taken in order to ensure that the acquired data is unbiased: we first make sure that the bifurcation jump occurs within a single point of the acquisition trace (we thus have to lower the filtering time constant of the lock-in compared to Fig. \ref{fig_lines}). Typically, we take one point every 40$~$ms with a frequency resolution typically 10 times smaller than the measured Gaussian spread. Second, we verify that we do not suffer from Brownian-type motion amplitude noise (at the mode frequency) that would activate relaxation of states when we are close enough to the bifurcation points \cite{AldridgeCleland,martial}. 
Such activated bifurcation generates non-Gaussian and asymmetric statistics, which is ramping-speed dependent \cite{AldridgeCleland}. No such characteristics have been seen in our experiments: we first check that the ramping speed (of order $0.1-1~$Hz/sec) does not change the measured histogram; and second, we add a controlled amount of force noise (at the mechanical resonance) in order to see when relaxation is indeed noise-activated \cite{martial}. We see that a force noise equivalent to a bath temperature of about $10^{6}~$K has to be reached in order to affect the frequency statistics.
Note that $10^{6}~$K is also the range of effective temperatures that are needed in order to see (asymmetric) frequency fluctuations transduced from Brownian motion through the Duffing nonlinearity \cite{PRBBrownian}.
Clearly, at 4.2$~$K with no added noise no such phenomena can occur. In the following, we make sure that no extra force noise is injected in the setup while measuring frequency fluctuations. 
Finally, the frequency drifts of our generators are characterized: we take two of them of the same brand, and measure the frequency stability of one against the clock of the other. Slow frequency fluctuations occur at the level of 1$~$mHz for 1$~$MHz signals over minutes, and 10$~$mHz for 10$~$MHz. 
This is at least two orders of magnitudes smaller than what is seen here over the same periods of time, and can be safely discarded.

We see that frequency fluctuations display a typical $1/f$-type behaviour (right inset in Fig. \ref{fig_lowstats} A), as reported by others \cite{hentz,tang}. 
Indeed, the time-trace has clearly some slowly drifting component (left inset in the same graph). This means that the statistics obtained {\it depends on the acquisition bandwidth}.
For pure $1/f$ noise, the standard deviation $\sigma_{f}$ (which is the square root of the power spectrum integral) depends on $\sqrt{\ln \left( f_{high}/f_{low}\right)}$, with $f_{high}$ the fastest frequency probed (defined from the time needed to acquire 1 bifurcation trace $\Delta t_{min}$, about 10 seconds) and $f_{low}$ the lower frequency cutoff (set by the total acquisition time, about 3 hours).
In order to be as quantitative as possible, we look for an estimate of frequency noise which is as much independent from the protocol as possible. We therefore study the {\it frequency jumps} $\delta f(t)=f_{down}(t_{i+1})-f_{down}(t_{i})$ instead of $f_{down}(t)$, see Fig. \ref{fig_lowstats} B left inset. The variance of this quantity $\sigma_{\delta \! f}^2$ is the well known Allan variance\cite{allan}, computed for acquisition time $\Delta t_{min}=t_{i+1}-t_{i}$. 
Note that this quantity safely suppresses equipment low frequency drifts (like, besides the one characterized for the generator, {\it e.g.} temperature drifts due to the $^4$He bath). 

For a perfectly $1/f$ noise, the Allan variance at $\Delta t_{min} \rightarrow 0$ is independent\cite{clelandnoise} of $f_{high},f_{low}$ (see Methods). However, our power spectrum fits $S_0/f^{1+\epsilon}$ with $\epsilon \approx 0.4 \pm 0.2$.   
We calculate that over the most extreme settings that have been used, our Allan variance should not have changed by more than 50$~$\% (and in the data presented here by much less). Moreover, to prove that the $\Delta t_{min} \rightarrow 0$ assumption can be applied we display in the right inset of Fig. \ref{fig_lowstats} B the power spectrum $S_{\delta \! f}(f)$ of $\delta f$ fluctuations: the data match the spectrum $\propto f$ directly computed from the fit of $S_{f}(f)$. 

We can then compare the values of $\sigma_f$ and $\sigma_{\delta \! f}$ that have been obtained. According to theory (see Methods) the first one should be about twice the second one in our conditions. This is indeed what we see in Fig. \ref{fig_lowstats}, confirming that $\sigma_{\delta f}$ is a good quantitative measurement of frequency fluctuations.
Besides, the frequency stability defined as $\sigma_{\delta f}/f_0 \approx 0.3~$ppm falls within the expected range according to reported measured devices \cite{hentz}. 
We thus demonstrate that our simple technique is functional: the key being that since the spectrum is $1/f$-type, we {\it do not need to be especially fast} to characterize frequency noise. 
With a Phase Locked Loop (PLL) setup one could measure fluctuations on much shorter timescales\cite{PLL}, but our 10$~$s repetition rate is perfectly well adapted. 

					\begin{figure}[h!]
			 \includegraphics[width=10cm]{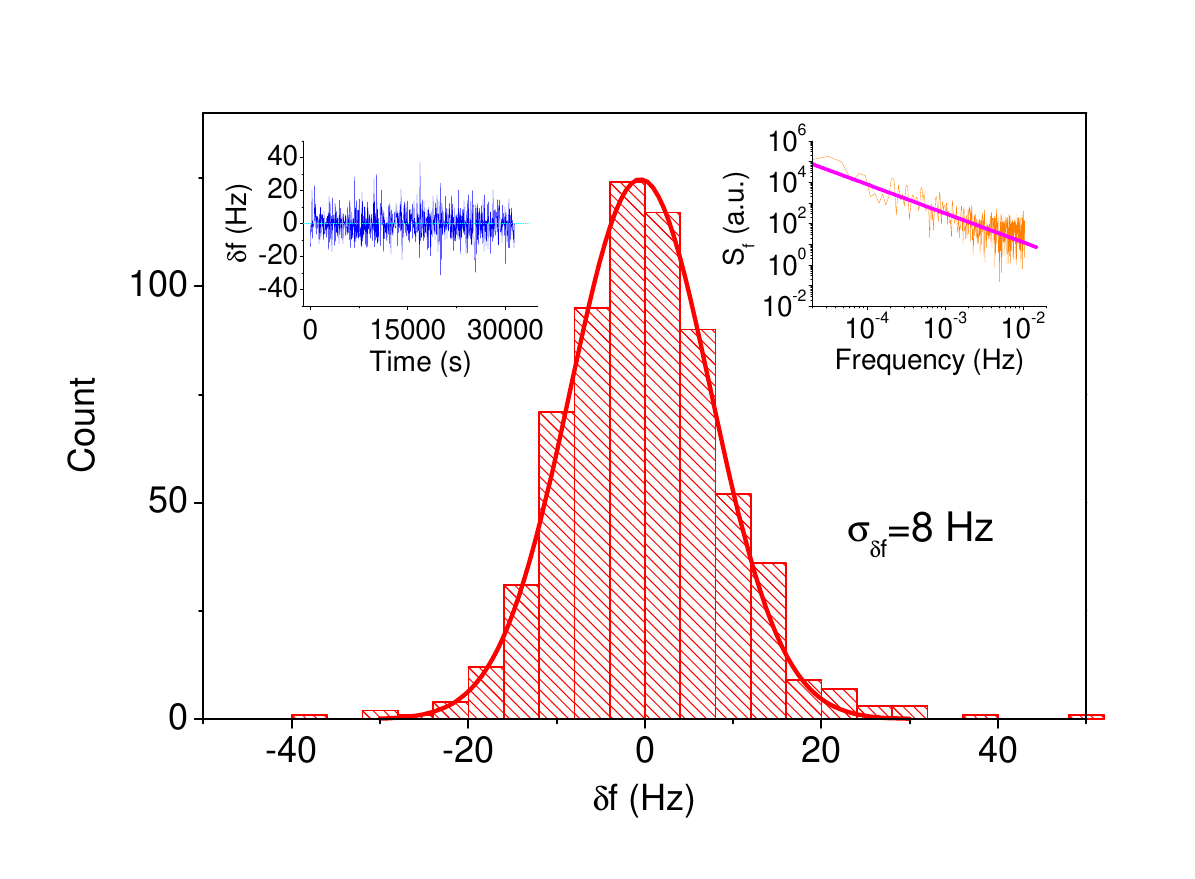}	
			\caption{\small{Large amplitude statistics. 
			Similar graphs as in Fig. \ref{fig_lowstats} obtained on the upper bifurcation point $f_{up}$ (frequency jumps time-trace in the left inset, power spectrum of frequency in the right inset and histogram of $\delta f$ in the main of the graph). The line is a Gaussian fit, while the power spectrum follows $1/f^{1.4}$ (see text).}}
			\label{fig_largestats}
		\end{figure}	
		
We now build on our 
method to thoroughly characterize frequency noise in SiN string devices.
Let us apply this technique to the upper bifurcation point $f_{up}$. 
The method explained above is easily reversed in sweep direction and threshold detection. 
Similar time-trace, spectrum and histogram to Fig. \ref{fig_lowstats} obtained this way are shown in Fig. \ref{fig_largestats}. 
We see that the power spectrum displays the same $1/f^{1+\epsilon}$ law with $\epsilon \approx 0.4$. This is true for the complete study we realized on the same device, and proves that different data sets can be consistently compared. The histogram is again Gaussian. But surprisingly, $\sigma_{\delta f}$ is now {\it much bigger} on the upper branch than on the lower one.
We therefore make a complete study as a function of driving force ({\it i.e.} motion amplitude). We discover that the standard deviation $\sigma_{\delta f}$ depends quadratically on	motion amplitude $R$. Measuring at another magnetic field $B_0$, we find that it also depends linearly on the $Q$ of the mechanical mode. 
However, measuring at different temperatures $T_0$ in the 1.4$~$K - 30$~$K range, we realize that the small amplitude value obtained is temperature-dependent, while the large amplitude one is not.

\begin{figure}[h!]
			 \includegraphics[width=11cm]{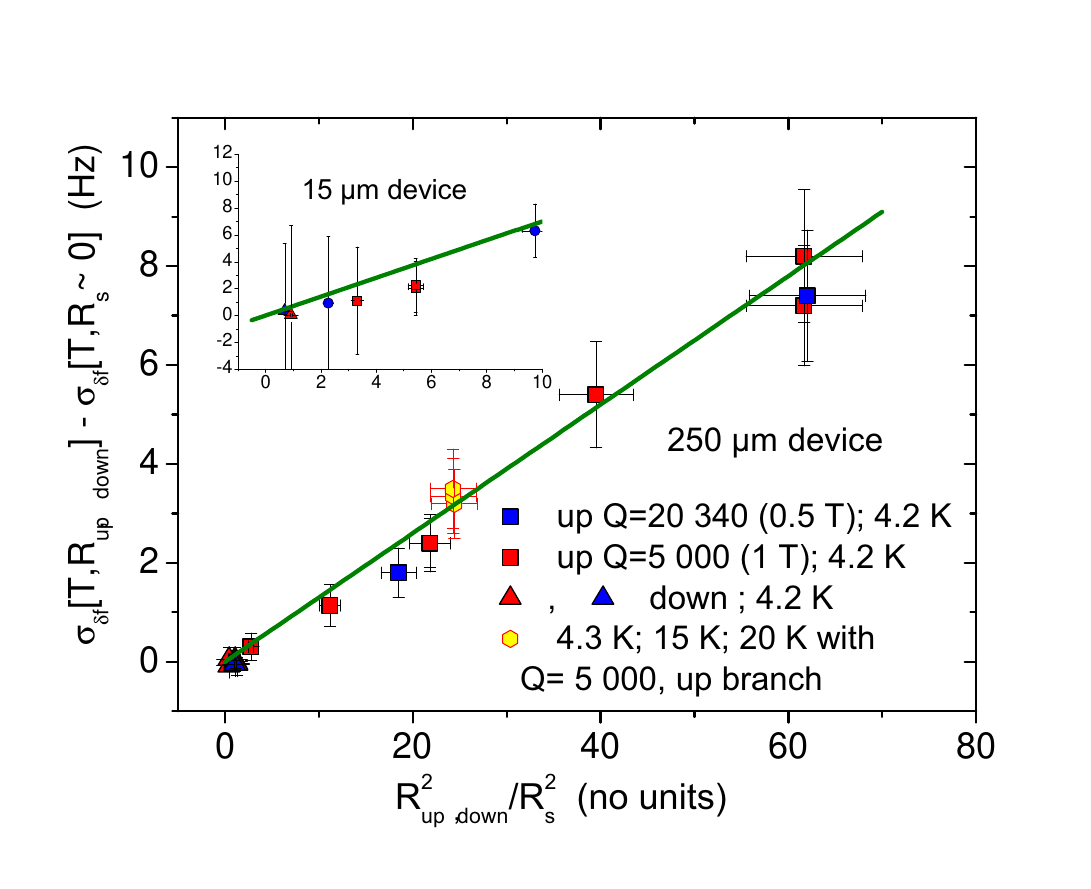}	
			\caption{\small{Universal plot for mode 1. 
			Frequency noise increase {\it vs} squared amplitude normalized to the spinodal value $R_s$, for the first mode $n=1$ of the 250$~\mu$m device. Various $T_0$ and $B_0$ (hence $Q$) have been used (see legend). Squares stand for upper branch bifurcation, triangles for lower. Inset: same result obtained with a 15$~\mu$m beam having $\beta = 1.1 \times 10^{19}~$Hz/m$^2$ 
(4.2$~$K and $Q=17\,000$ red squares; 800$~$mK and $Q=31\,000$ blue dots; the magnetomotive field broadening was negligible). The green line is a linear fit (see text).}}
			\label{fig_univ1}
		\end{figure}	

This suggests the normalized plot of Fig. \ref{fig_univ1}, where the increase $\sigma_{\delta f}(T,R_{up,down})$ from the extrapolated $\sigma_{\delta f}(T,R_{s})$ is plotted against the normalized variable $R_{up,down}^2/R_s^2$. The notation $R_{s}\sim 0$ means that the value is obtained from the linear fit, extrapolating at $R \rightarrow 0$. 
In order to verify the robustness of the result, we realize the same analysis with a similar device of 15$~\mu$m length. Some typical data is displayed in the inset of Fig. \ref{fig_univ1}. 
The noise properties obtained for this other device are very similar to the initial 250$~\mu$m long beam (but the quadratic dependence is different). However the spectra better fit with $\epsilon \approx 0.2 \pm 0.2$.

\begin{figure}[h!]
			 \includegraphics[width=11cm]{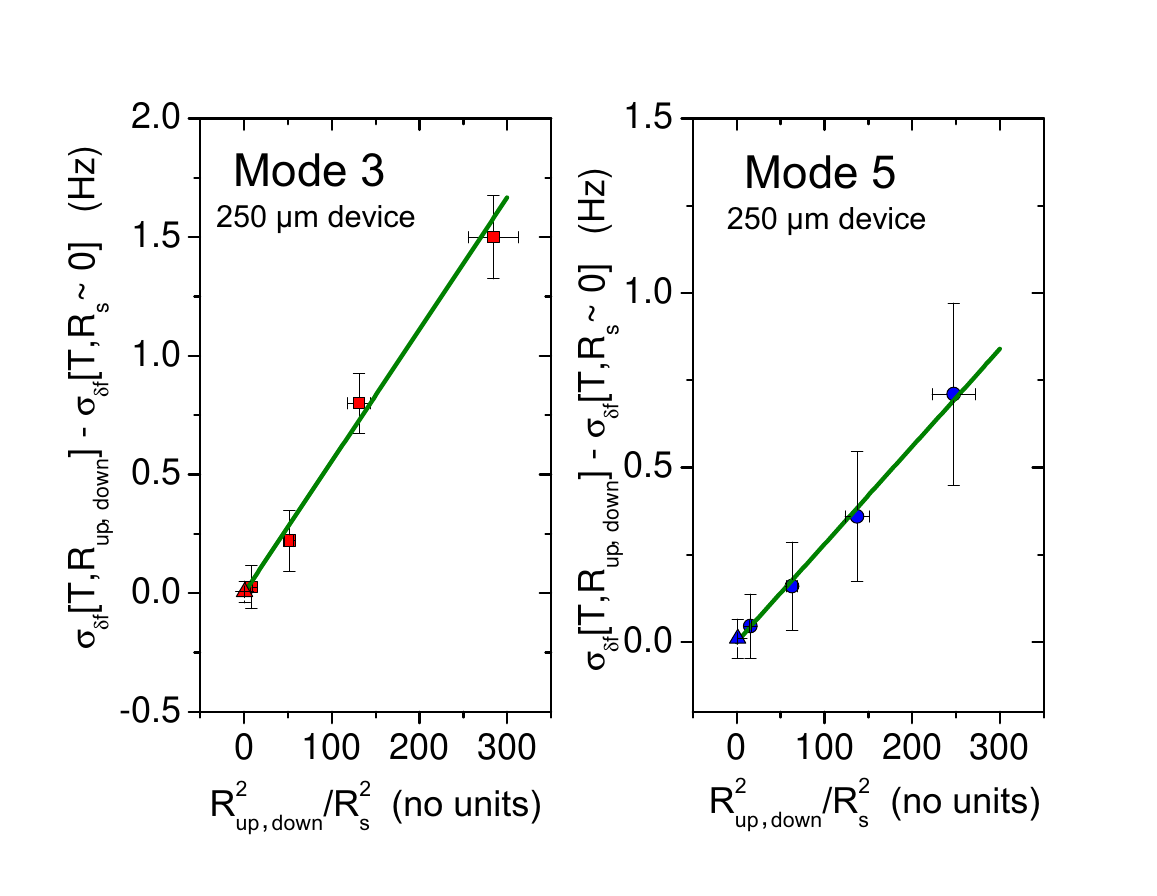}	
			\caption{\small{Universal plot for modes 3 and 5. 
			Normalized frequency noise plot for the same $250~\mu$m device, for modes $n=3$ (left) and $n=5$ (right). The lines are linear fits (see text).}}
			\label{fig_univ2}
		\end{figure}

We proceed with similar measurements performed on modes $n=3$ and $n=5$ of the 250$~\mu$m beam sample. All spectra display the same $1/f^{1.4}$ dependence as the $n=1$ mode.
The same normalized plots are displayed in Fig. \ref{fig_univ2}. 
However, this time the inferred quadratic dependencies are much weaker. 

The nonlinear dependence of the frequency noise is rather unexpected; indeed, the nonlinear dephasing features observed for carbon-based devices \cite{venstra} have not been reported for nitride structures \cite{NJPus}.
A possible source for such an effect could be a {\it purely intrinsic} property of the bifurcation effect itself. However, since our statistics could not be altered by  reasonable changes in effective temperatures and frequency-sweep ramping speeds, such an explanation is improbable.
If we then suppose the bifurcation process to be perfectly ideal, the nonlinear frequency noise observed should originate in {\it one of the parameters} defining the bifurcation frequencies. When the experiment is performed reasonably far from the spinodal point (which is always our case), we have:
\begin{eqnarray}
f_{up} & \approx & f_0 + \delta \phi(t) + \beta R_{up}^2 , \label{simple1} \\
f_{down}  & \approx & f_0 + \delta \phi(t) + 3 \beta R_{down}^2 , \\
R_{up} & \approx  & R_{max} , \\
R_{down} & \approx & \frac{1}{2^{2/3}}\left( \frac{R_{max} \, \Delta f }{\beta}\right)^{1/3} ,\label{simple4}
\end{eqnarray}
adapted from Eqs. (\ref{fup}-\ref{fdown}), where we explicitly introduced the stochastic frequency variable $\delta \phi(t)$.
For strings\cite{lifshitz} $\beta \propto 1/(m_0 f_0) \left( \frac{E_Y \,A}{L^3} \right)$, the nonlinear frequency noise could be caused by (Gaussian) fluctuations of the Young's modulus $E_Y=E_0+\delta E(t)$. However, to have the measured characteristics, this noise would have to be $\delta E/E_0 \propto 1/\Delta f$ and mode-dependent, which is difficult to justify: this explanation seems again unphysical.

The only possibility left is an {\it internal motional noise} with $R_{max}=R_{max_0}+\delta R$, leading to fluctuations $\propto \left( \frac{\delta R}{R_{max_0}} \right) \beta R_{max_0}^2 $. The proper scalings, as reported in Figs. \ref{fig_univ1} and \ref{fig_univ2}, are thus only achieved by assuming {\it damping noise} $\delta \Gamma(t)$ with $\frac{\delta R}{R_{max_0}} = - \frac{\delta \Gamma}{\Delta f}$. 
As a result, it follows from Eqs. (\ref{simple1}-\ref{simple4}): 
$f_{up} (t) \approx  f_0 + \delta \phi(t) + \beta R_{max_0}^2 \left(1- 2 \frac{\delta \Gamma(t)}{\Delta f}\right)$, $f_{down} (t)  \approx  f_0 + \delta \phi(t) + \frac{3}{2^{4/3}} (\Delta f^2 \, \beta R_{max_0}^2)^{1/3}$ together with $f_s (t) = f_0 +\delta \phi(t) + \frac{\sqrt{3}}{2} \Delta f \left(1+  \frac{\delta \Gamma(t)}{\Delta f}\right)$.
This means that both bifurcation frequencies suffer from frequency noise $\delta \phi$, while {\it only the upper branch} experiences damping fluctuations $\delta \Gamma$: they are {\it amplified} by the measurement method through a factor $\beta R_{max_0}^2/\Delta f$. 
Note that the frequency noise extrapolated at $R \rightarrow 0$ on the upper branch matches the one of the lower branch, but does not equal the one obtained at the spinodal point, simply because the expressions Eqs. (\ref{simple1}-\ref{simple4}) do not apply near $R_s$; this is emphasized through the writing $R_s \sim 0$ in our graphs.

\begin{table}[h!]
\begin{center}
\hspace*{-2cm}
\begin{tabular}{|c|c|c|c|c|c|}    \hline
 Mode number $n$ & Freq. $f_0$ & Unloaded $Q$ & Duffing $\beta $  & $\sigma_{\delta \! f}$ at $R_s \sim 0$  & $\sigma_{\delta \Delta\! f}$ from $R_{up} \gg R_s$ \\   \hline \hline
$n=1$    & $0.905~$MHz       & $600\,000 \pm 10~\%$         & $8.5 \pm 0.5 \times 10^{15}~$Hz/m$^2$    &   $0.28\pm 0.05~$Hz  & $0.11~$Hz $\pm 10~\%$ \\    \hline
$n=3$    & $2.68~$MHz        & $450\,000 \pm 10~\%$         & $1.25 \pm 0.2 \times 10^{17}~$Hz/m$^2$   &   $0.1\pm 0.02~$Hz   & $0.005~$Hz $\pm 15~\%$ \\    \hline
$n=5$    & $4.45~$MHz        & $400\,000 \pm 20~\%$         & $5.7 \pm 0.5 \times 10^{17}~$Hz/m$^2$    &   $0.09\pm 0.02~$Hz  & $0.0025~$Hz $\pm 30~\%$ \\    \hline
\end{tabular}
\caption{\label{ValuesModes} Mode parameters, frequency and damping fluctuations for modes $n=1, 3$ and $5$ (250$~\mu$m long beam, 4.2$~$K).}
\end{center}
\end{table}

\begin{figure}[h!]
			 \includegraphics[width=11cm]{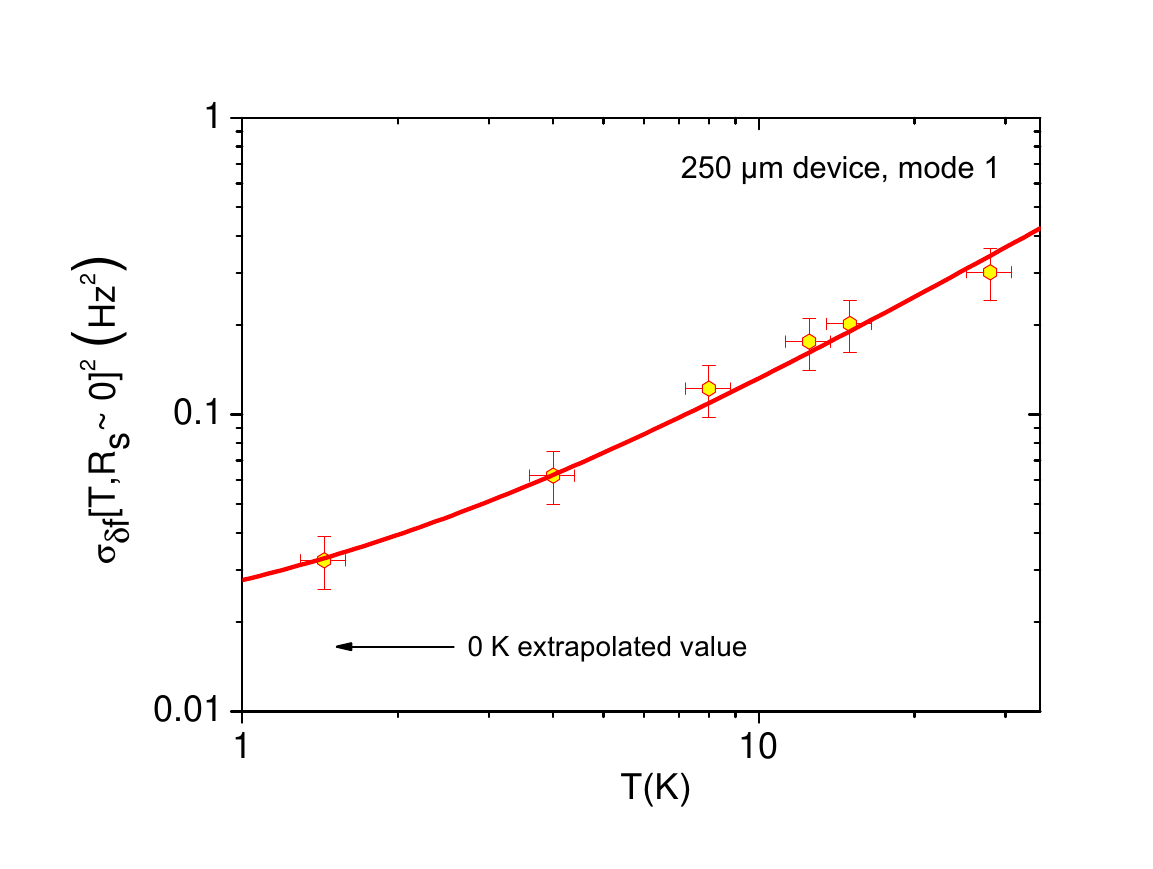}	
			\caption{\small{Temperature dependence of frequency fluctuations. 
			Allan variance $\sigma_{\delta f}^2$ as a function of temperature (first mode of 250$~\mu$m device), as obtained for small motion $R\rightarrow R_s \sim 0$. The line is a linear fit, with the $T_0=0~$K extrapolated value emphasized by the arrow (see text). The variance of damping fluctuations $\sigma_{\delta \Delta f}^2$ is constant in the same range of temperatures. 
			}}
			\label{fig_Tdep}
		\end{figure}

The Allan deviation $\sigma_{\delta f}$ extrapolated to $R \rightarrow 0$ is thus characteristic of the frequency noise $\delta \phi$, while the slopes of the graphs in Figs. \ref{fig_univ1} and \ref{fig_univ2} are $\frac{2}{\sqrt{3}}$ times the Allan deviation $\sigma_{\delta \Delta f}$ of the damping fluctuations. To our knowledge, the latter has not been reported in the literature so far.
The mode parameters together with these 4.2$~$K frequency and damping noise figures are summarized in Table \ref{ValuesModes}.
$\sigma_{\delta \Delta f}$ is temperature-independent in the range studied, while $\sigma_{\delta f}^2$ is linear in $T_0$; this is displayed in Fig. \ref{fig_Tdep}. 
The same temperature-dependence of frequency noise has been reported in Ref. \cite{tang} (within an overall scaling factor) for a very similar device. In order to compare the various results, values from the literature are presented in Table \ref{ValuesDevice}. 
We give the Allan deviation when it is reported, otherwise we list the direct frequency noise; the damping noise in the third line is recalculated from Fig. IV.19 in Ref. \cite{martialthesis}.

\begin{table}[h!]
\begin{center}
\hspace*{-2cm}
\begin{tabular}{|c|c|c|c|c|}    \hline
 Device & Freq. $f_0$ & $Q $  & $\sigma_{\delta \! f}$ or $\sigma_{f_0}$ & $\sigma_{\delta \Delta\! f}$ \\   \hline \hline
$250~\mu$m SiN/Al d.c. beam 4.2$~$K (this work)    & $0.905~$MHz       & $600\,000 \pm 10~\%$      &   $0.28\pm 0.05~$Hz & $0.11~$Hz $\pm 10~\%$  \\    \hline
$15~\mu$m SiN/Al d.c. beam 4.2$~$K(this work)    & $17.5~$MHz       & $18\,000 \pm 10~\%$      &   $1.45\pm 0.1~$Hz  & $0.6~$Hz $\pm 10~\%$ \\    \hline
$2 \times 3~\mu$m Si/Al goalpost 4.2$~$K Refs. \cite{martial,martialthesis}    & $7.1~$MHz       & $4\,700 \pm 10~\%$      &   $1\pm 0.5~$Hz on $f_0$ & $0.35~$Hz $\pm 10~\%$ \\    \hline
$380~\mu$m SiN d.c. beam 5$~$K Ref. \cite{tang}    & $0.64~$MHz       & $2\,200\,000$      &   $0.01~$Hz on $f_0$ & X \\    \hline
$3.2~\mu$m Si cantilever Room Temp. Ref. \cite{hentz}    & $45.2~$MHz       & $6\,000 $      &   $36~$Hz  & X \\    \hline
\end{tabular}
\caption{\label{ValuesDevice} Mode parameters, frequency and damping fluctuations for different devices (fundamental flexure $n=1$). The damping noise figure in the third line is recalculated from Fig. IV.19 in Ref. \cite{martialthesis}.}
\end{center}
\end{table}

We can only speculate on the microscopic mechanisms behind the reported features. 
The entities generating such noise are thought to be atomic-scale two level systems (TLS), which could be defects or intrinsic to the materials in use \cite{tang, clelandnoise, tlspaper, martialthesis}. A signature that also supports this view is the presence of {\it telegraph frequency noise} in many NEMS experiments (see {\it e.g.} Ref. \cite{martialthesis}). Ref. \cite{clelandnoise} analyzed frequency noise in beams, {\it i.e.} structures with no built-in stress. These Authors assumed that thermally activated motion of a defect in a double well potential from one minimum to the other caused a shift in the local Young's modulus. The motion of many such defects 
(following the mathematical arguments of Ref. \cite{dutta}) 
then causes a change in the average Young's modulus and consequently a change in the resonance frequency of the beam
with power spectrum $\propto T/f$. 
The same argument was applied in Ref. \cite{tang} to analyze the frequency noise of string structures, where the built-in stress is large, even though the resonance frequency is nearly independent of the Young's modulus in the high-stress limit. We thus believe that it is more appropriate, in the interpretation of the present measurements on strings, to consider frequency fluctuations due to {\it stress fluctuations}.
Indeed, point defects in crystals are characterized as elastic dipoles that cause an orientation dependent change in the strain (and consequently the stress) of a crystal \cite{nowick}.


In our highest $Q$ device, for the first flexure $n=1$ frequency fluctuations at 4.2$~$K represent about 20$~$\% of the linewidth, and damping fluctuations about 5$~$\%. These parameters fall with mode number $n$ (see Tab. \ref{ValuesModes}), while both frequency $f_0$ and linewidth $\Delta f$ increase. 
This means that the effect of fluctuations is {\it the strongest} on the first mode, but is usually difficult to visualize on standard frequency-sweeps or time-decay data; it is for instance expected that for a device rather equivalent to our 250$~\mu$m beam, the measurements performed in Ref. \cite{NJPus} did not report any such features (see Methods).
		
\section{Conclusion}

In this Letter we present a very simple and reliable method to measure and characterize frequency noise in bistable resonators.
The technique has been employed to describe thoroughly the intrinsic frequency noise of high-stress silicon-nitride doubly-clamped beams. The measurements have been performed at low temperatures in cryogenic vacuum, on two devices of very different lengths/fundamental resonance frequencies. 
The 3 first symmetric flexural modes of the longest beam have been studied. 

We report on the Allan deviations of the frequency noise, presenting the same basic features as in Refs. \cite{hentz,tang}: spectra of $1/f$-type, scaling linearly with temperature. The reported magnitudes of the noise $\delta f/f_0$ fall in the  $0.5 - 0.01~$ppm range, as expected for MHz devices. 

We have also found unexpectedly {\it damping fluctuations}, which are amplified in the vicinity of bifurcation points. Our technique 
seems the most adapted for the detection and the quantification of such a noise process to date. 
We find that damping noise can be as large as about 5$~$\% of the total width of the resonance peak in our highest $Q$ devices. It also sets a finite resolution attainable for the measurement of the frequency position of bifurcation points.

These features seem ubiquitous to all NEMS devices, and we do believe that damping noise and frequency noise originate in the same microscopic mechanism. But the latter remains elusive, and the most discussed candidate is based on Two-Level 
Systems (TLS) \cite{clelandnoise,tang}. 
Because of the strength of the frequency noises reported here for high-stress devices, we propose that TLS are responsible for noise on the {\it stored stress} in the structure instead of the Young's modulus, as it was proposed in earlier papers discussing stress-free beams ({\it e.g.} Ref. \cite{clelandnoise}).

Our technique can be easily adapted to any types of devices, including MoS$_2$ and carbon-based systems in which nonlinear frequency noise has been reported \cite{venstra}. We think that it should help advance the understanding of the underlying fundamental microscopic mechanisms that also significantly degrade the properties of existing NEMS devices, and hinder their applicability. 
	
\section{Methods}

\subsection*{Device Fabrication}
\small{The structure was fabricated using e-beam lithography on a silicon substrate covered with a $100~\nano\meter$ silicon nitride (SiN) layer. The stochiometric nitride was grown using low pressure chemical
vapor deposition (LPCVD) at the Cornell NanoScale Science \& Technology Facility (CNF). 
It stores a biaxial stress of about 1$~$GPa. A 30$~$nm Aluminum coating has been evaporated onto the sample in a Plassys e-gun machine. Its resistance at low temperature is about 1$~$k$\Omega$ for the 250$~\mu$m long device, and 100$~\Omega$ for the 15$~\mu$m one.
It served as a mask for the structure during the SF$_6$ Reactive Ion Etching (RIE) step used to pattern the SiN. The structure was then released using a final $\mathrm{XeF_2}$ etching of the underlying silicon.}
\subsection*{Measurements}
\small{The voltage drive was delivered by a Tektronix AFG3252 generator, through a $1~\kilo\ohm$ bias resistance which created the drive current. The motion was actuated with the magnetomotive technique through a force $F(t) = I_0 \zeta L B_0 \, \cos(2 \pi f \, t)$, which also leads to the detection of the velocity $\dot{x}(t)$ of the oscillation through a voltage $V(t)=\zeta L B_0 \, \dot{x}(t)$. $\zeta$ is a mode-dependent shape factor \cite{clelandsensors}. In the high $Q$ limit, the velocity in frequency domain is $i \omega_0 \, x(\omega)$ with $\omega_0=2 \pi f_0$, hence an inverted definition for the signal quadratures $X$ and $Y$ with respect to displacement $x(t)$. 
Due to the symmetry of the scheme, only symmetric modes ($n= 1, 3, 5 \cdots$) can be addressed ($\zeta =0$ otherwise).
The magnetic field was generated with a small superconducting coil fed with a 10 A Kepco current source.
The detected signal was processed with a Stanford SR 844 RF lock-in amplifier. 
Due to the finite impedance of the electric circuit seen by the NEMS (whose own characteristic impedance varies as $B_0^2$), the mechanical resonances are loaded by an additional damping $\propto B_0^2$. This provides the ability to tune the quality factors {\it in situ} \cite{clelandsensors}. 
Our calibration procedure is described in Ref. \cite{RSI}. It enables us to give all mechanical parameters in S.I. units (we thus quote $X$,$Y$ in meters), while minimizing the loading effect. Loading is negligible for the 15$~\mu$m beam, but still large in the 250$~\mu$m device.
\subsection*{Spectra Mathematical Properties}
Let us consider a frequency power spectrum for the stochastic resonance frequency $\omega_0=2 \pi f_0$ of type $S_{\omega_0} (\omega) = A_0/ \left| \omega \right|^{1+\epsilon}$ (defined from $-\infty$ to $+\infty$). The variance can be defined from the integral of the spectrum, leading to $\sigma_{\omega_0}^2=\frac{1}{2 \pi} \int S_{\omega_0} d \omega = \frac{2 A_0}{2 \pi} \left[ \frac{\omega_{low}^{-\epsilon}-\omega_{high}^{-\epsilon}}{\epsilon} \right]$ with $\omega_{low}$ and $\omega_{high}$ the lower and higher frequency cutoffs imposed by the experiment. For $\epsilon \rightarrow 0$ we have $\sigma_{\omega_0}  \propto \sqrt{\ln \left[ \omega_{high}/ \omega_{low}\right]}$. Since $\delta \omega = \omega_0(t_{i+1}) -\omega_0(t_{i}) \approx \partial \omega_0(t) / \partial t  \times \Delta t_{min}$ when $\Delta t_{min} \rightarrow 0$, we have $S_{\delta \omega} (\omega) \approx \Delta t_{min}^2 \, \omega^2  \, S_{\omega} (\omega)$. Thus $\sigma_{\delta \omega}^2 \approx \frac{2 A_0}{2 \pi} \frac{\left( \Delta t_{min}  \, \omega_{high}\right)^2}{2} \left[ \omega_{high}^{-\epsilon} \frac{1-\left( \omega_{low}/\omega_{high} \right)^{2-\epsilon}}{1-\epsilon/2}   \right]$. The Fourier Transform imposes $\Delta t_{min}  \, \omega_{high} \approx \pi$, and in the case $\epsilon \rightarrow 0$ we obtain $\sigma_{\delta \omega} \propto \sqrt{1-\left( \omega_{low}/ \omega_{high}\right)^2}$, which is essentially independent of the cutoffs \cite{clelandnoise}. For $\epsilon \neq 0$, a small dependence to the bandwidth appears in the Allan variance $\sigma_{\delta \omega}^2$. For our acquisition bandwidths, this does not result in a too large scatter in data (within error bars). 
\subsection*{Impact on Frequency-domain \& Time-domain Measures}
With a noise of type $S_{\omega_0} (\omega) = A_0/ \left| \omega \right|^{1+\epsilon}$, we can take as an estimate of the relevant fluctuations timescale $\tau_c^{-1} \sim \omega_{low}/\pi$: the weight is at the lowest accessible frequencies.
We thus always verify $\sigma_{\omega_0} \tau_c \gg 1$, which means that the phase diffusion of the mechanical mode is in the Inhomogeneous Broadening limit (IB), in analogy with Nuclear Magnetic Resonance \cite{zhanganddyk,PRBBrownian,NJPus}. In frequency-domain, the response $\chi_{meas}(\omega)$ is the convolution of the standard (complex-valued, defining the two quadratures) susceptibility $\chi(\omega)$ with the (Gaussian) distribution of frequencies $p(\delta \varphi) = 1/\sqrt{2 \pi \sigma_{\omega_0}^2 } \exp \left(-\frac{1}{2} \delta \varphi^2/\sigma_{\omega_0}^2 \right)$, with $\delta \varphi = 2 \pi \delta \phi$ in Rad/s. 
This means that at each scanned frequency $\omega$, the measurement is performed over a long enough timescale such that all fluctuations are explored.
On the other hand, the small damping fluctuations are simply filtered out by the acquisition setup (here, a lock-in amplifier): they have no relevant impact on the resonance peak measured, even at very large motion amplitudes. 
We conclude that only frequency noise will contribute to the definition of a $T_2$, the decoherence time involving relaxation $T_1$ and dephasing $\sigma_{\omega_0}$ \cite{NJPus}.
In time-domain, in the linear regime the complex susceptibility $\bar{\chi}_{meas}(t)$ ({\it i.e.} decay of the two quadratures) is simply the Fourier Transform (FT) of $\chi_{meas}(\omega)$. It can also be written $\bar{\chi}_{meas}(t) = \left\langle \exp \left(i \delta \varphi t \right) \right\rangle \bar{\chi}(t)$ with $\bar{\chi}(t)$ the FT of $\chi(\omega)$ and $\left\langle \exp \left(i \delta \varphi t \right) \right\rangle = \exp \left(-\frac{1}{2} \sigma_{\omega_0}^2 t^2 \right)$ the average over frequency fluctuations.
In the nonlinear regime, the averaged decay of the two quadratures writes $\bar{\chi}_{meas}(t) = \left\langle \exp \left(i \delta \varphi t \right) \right\rangle \left\langle \exp \left(-i 2 \frac{\beta R_{max_0}^2}{\Delta \omega} \delta \gamma \, t \,\kappa\left[t\right] \right) \right\rangle \bar{\chi}(t)$ with $\kappa\left[t\right]=\frac{1-\exp \left[ -t \Delta \omega \right]}{t \Delta \omega}$ and $\bar{\chi}(t)$ defined in Ref. \cite{PRBus}, the second average being over damping fluctuations $\delta \gamma = 2\pi \delta \Gamma$; we wrote $\Delta \omega = 2 \pi \Delta f$ in Rad/s.  
The function $\kappa$ is characteristic of the decay of the nonlinear frequency pulling due to the Duffing term\cite{PRBus}, $\propto R_{max_0}^2$. In practice, this assumes that both quadratures are measured independently, averaging many decay traces starting from the same (noisy) $t=0$ amplitude $R_{max}$, imprinted by the slow fluctuations of damping $\delta \gamma$.
The second average can be explicitly calculated: $\left\langle \exp \left(-i 2 \frac{\beta R_{max_0}^2}{\Delta \omega} \delta \gamma \, t\, \kappa\left[t\right]  \right) \right\rangle = \exp \left[ -\frac{1}{2} \left(2 \frac{\beta R_{max_0}^2}{\Delta \omega}\right)^2 \sigma_{\delta \Delta \omega}^2 \left(  t \,\kappa\left[t\right]\right)^2 \right]$. 
In time-domain, the impact of damping fluctuations (of variance $\sigma_{\delta \Delta \omega}^2$) is thus {\it amplified} by the same term as in the bifurcation measurement: $\frac{\beta R_{max_0}^2}{\Delta \omega}$. However, to have a measurable effect (within experimental error bars) both amplitude $R_{max_0}$ and fluctuations $\sigma_{\delta \Delta \omega}$ have to be very large; in experiments of the type of Ref. \cite{NJPus} based on devices similar to our 250$~\mu$m, no such effect has been reported.
Measuring the decay of the two quadratures leads then to the definition of $\bar{T}_2$, roughly equivalent to $T_2$ \cite{NJPus}.
Note that the decay of the motion amplitude $\left|\bar{\chi}(t)\right|^2$ remains unaffected by frequency and damping noise, leading to the proper $T_1$ definition \cite{NJPus}.  


%
%


\begin{acknowledgement}

We thank J. Minet and C. Guttin for help in setting up the experiment, as well as J.-F. Motte, S. Dufresnes and T. Crozes from facility Nanofab for help in the device fabrication. We acknowledge support from the ANR grant MajoranaPRO No. ANR-13-BS04-0009-01 and the ERC CoG grant ULT-NEMS No. 647917. This work has been performed in the framework of the European Microkelvin Platform (EMP) collaboration. At Cornell we acknowledge support from the NSF under DMR 1708341.

\end{acknowledgement}


\begin{thebibliography}
\small{

\bibitem{moser} Moser, J.; G\"uttinger, J.; Eichler, A.; Esplandiu, M. J.; Liu, D. E.; Dykman, M. I.;  Bachtold, A. Ultrasensitive Force Detection with a Nanotube Mechanical Resonator. {\it Nat. Nanotechnol.} {\bf 2013}, 8, 493-496.

\bibitem{liroukes} Li, M.; Tang, H. X.; Roukes, M. L. Ultra-sensitive NEMS-based Cantilevers for Sensing, Scanned Probe and Very High-frequency Applications. {\it Nat. Nanotechnol.} {\bf 2007}, 2, 114-120.

\bibitem{rugar} Rugar, D.; Budakian, R.; Mamin, H. J.; Chui, B. W. Single Spin Detection by Magnetic Resonance Force Microscopy. {\it Nature} {\bf 2004}, 430, 329-332.

\bibitem{chaste}  Chaste, J.; Eichler, A.;  Moser, J.;  Ceballos, G.; Rurali, R.; Bachtold, A. A Nanomechanical Mass Sensor with Yoctogram Resolution. {\it  Nat. Nanotechnol.} {\bf 2012}, 7, 301-304.

\bibitem{quantum} Pirkkalainen, J.-M.; Cho, S. U.; Li, J.; Paraoanu, G. S., Hakonen, P. J.; Sillanp\"a\"a, M. A. Hybrid Circuit Cavity Quantum Electrodynamics with a Micromechanical Resonator. {\it Nature} {\bf 2013}, 494, 211-215.

\bibitem{clelandnoise} Cleland, A. N.; Roukes, M. L. Noise Processes in Nanomechanical Resonators. {\it J. Appl. Phys.} {\bf 2002}, 92, 2758-2769.

\bibitem{hentz} Sansa, M.; Sage, E.; Bullard, E. C.; G\'ely, M.; Alava, T.; Colinet, E.; Naik, A. K.; Villanueva, L. G.; Duraffourg, L.; Roukes, M. L.;  Jourdan, G.; Hentz, S. Frequency Fluctuations in Silicon Nanoresonators. {\it Nat. Nanotechnol.} {\bf 2016}, 11, 552-558.

\bibitem{bachtolddykman} Zhang, Y.; Moser, J.; G\"uttinger, J.; Bachtold, A.; Dykman, M. I. Interplay of Driving and Frequency Noise in the Spectra of Vibrational Systems. {\it Phys. Rev. Lett.} {\bf 2014}, 113, 255502.

\bibitem{greywall} Greywall, D. S.; Yurke, B.; Busch, P. A.; Pargellis, A. N.; Willett, R. L. Evading Amplifier Noise in Nonlinear Oscillators. {\it Phys. Rev. Lett.} {\bf 1994}, 72, 2992-2995.

\bibitem{PLL} Feng, X. L.; He, R. R.; Yang, P. D.; Roukes, M. L. Phase Noise and Frequency Stability of Very-High Frequency Silicon Nanowire Nanomechanical Resonators. {\it Transducers Eurosens. XXVII, Int. Conf. Solid-State Sens., Actuators Microsyst., 17th.} {\bf 2007}, 327-330.

\bibitem{tang} Fong, K. Y.; Pernice, W. H. P.; Tang, H. X. Frequency and Phase Noise of Ultrahigh Q Silicon Nitride Nanomechanical Resonators. {\it Phys. Rev. B} {\bf 2012}, 85, 161410(R).

\bibitem{venstra} Schneider, B. H.; Singh, V.; Venstra, W. J.; Meerwaldt, H. B.; Steele, G. A. Observation of Decoherence in a Carbon Nanotube Mechanical Resonator. {\it  Nat. Commun.} {\bf 2014}, 5, 5819.

\bibitem{NJPus} Maillet, O.; Vavrek, F.; Fefferman, A. D.; Bourgeois, O.; Collin, E. Classical Decoherence in a Nanomechanical Resonator. {\it New J. Phys.} {\bf 2016}, 18, 073022.

\bibitem{atalayadykman} Atalaya, J.; Isacsson, A.; Dykman, M. I. Diffusion-induced Dephasing in Nanomechanical Resonators. {\it Phys. Rev. B} {\bf 2011}, 83, 045419.

\bibitem{diffuseRoukes} Yang, Y. T.; Callegari, C.; Feng, X. L.; Roukes, M. L. Surface Adsorbate Fluctuations and Noise in Nanoelectromechanical Systems. {\it Nano Lett.} {\bf 2011}, 11, 1753-1759.

\bibitem{zhanganddyk} Zhang, Y.; Dykman, M. I. Spectral Effects of Dispersive Mode Coupling in Driven Mesoscopic Systems. {\it Phys. Rev. B} {\bf 2015}, 92, 165419.

\bibitem{PRBBrownian} Maillet, O.; Zhou, X.; Gazizulin, R. R.; Maldonado Cid, A. I.; Defoort, M.; Bourgeois, O.; Collin, E. Nonlinear Frequency Transduction of Nanomechanical Brownian Motion. {\it Phys. Rev. B} {\bf 2017}, 96, 165434.

\bibitem{nonlindephase} Atalaya, J.; Kenny, T. W.; Roukes, M. L.; Dykman, M. I. Nonlinear Damping and Dephasing in Nanomechanical Systems. {\it Phys. Rev. B} {\bf 2016}, 94, 195440.

\bibitem{dutta} Dutta, P.; Horn, P. M. Low-frequency Fluctuations in Solids: 1/f Noise. {\it Rev. Mod. Phys.} {\bf 1981}, 53, 497-516.

\bibitem{tlspaper} Behunin, R. O.; Intravaia, F.; Rakich, P. T. Dimensional Transformation of Defect-induced Noise, Dissipation, and Nonlinearity. {\it Phys. Rev. B} {\bf 2016}, 93, 224110.

\bibitem{TLSnems1} Hoehne, F.;  Pashkin, Y. A.; Astafiev, O.; Faoro, L.; Ioffe, L. B.; Nakamura, Y.; Tsai, J. S. Damping in High-frequency Metallic Nanomechanical Resonators. {\it Phys. Rev. B} {\bf 2010}, 81, 184112.

\bibitem{TLSnems2} Venkatesan, A.; Lulla, K. J.; Patton, M. J.; Armour, A. D.; Mellor, C. J.; Owers-Bradley, J. R. Dissipation Due to Tunneling Two-level Systems in Gold Nanomechancial Resonators. {\it Phys. Rev. B} {\bf 2010}, 81, 073410.

\bibitem{graphene} Miao, T.; Yeom, S.; Wang, P.; Standley, B.; Bockrath, M. Graphene Nanoelectromechanical Systems as Stochastic-Frequency Oscillators. {\it Nano Lett.} {\bf 2014}, 14, 2982-2987.

\bibitem{clelandsensors} Cleland, A. N.; Roukes, M. L. External Control of Dissipation in a Nanometer-scale Radiofrequency Mechanical Resonator. {\it  Sens. Actuators} {\bf 1999} 72, 256-261.

\bibitem{RSI} Collin, E.; Defoort, M.; Lulla, K.; Moutonet, T.; Heron, J.-S.; Bourgeois, O.; Bunkov, Y. M.; Godfrin, H. {\it In situ} Comprehensive Calibration of a Tri-port Nano-electro-mechanical Device. {\it Rev. Sci. Instrum.} {\bf 2012}, 83, 045005.

\bibitem{lifshitz} Lifshitz, R.; Cross, M. C. Nonlinear Dynamics of Nanomechanical and Micromechanical Resonators. In {\it Reviews of Nonlinear Dynamics and Complexity}; Schuster, H.G. Ed.; Wiley: VCH, 2008; pp 1 - 52.

\bibitem{roukes} Matheny, M. H.; Villanueva, L. G.; Karabalin, R. B.; Sader, J. E.; Roukes, M. L. Nonlinear Mode-Coupling in Nanomechanical Systems. {\it Nano Lett.} {\bf 2013}, 13, 1622-1626. 

\bibitem{PRBus} Collin, E.; Bunkov, Y. M.; Godfrin, H. Addressing Geometric Nonlinearities with Cantilever Microelectromechanical Systems:
Beyond the Duffing Model. {\it Phys. Rev. B} {\bf 2010}, 82, 235416.

\bibitem{landau} Landau, L. D.; Lifshitz, E. M. In {\it Mechanics}; Elsevier: New York, 1976.

\bibitem{AldridgeCleland} Aldridge, J. S.; Cleland, A. N. Noise-enabled Precision Measurements of a Duffing Nanomechanical Resonator. {\it Phys. Rev. Lett.} {\bf 2005}, 94, 156403.

\bibitem{martial} Defoort, M.; Puller, V.; Bourgeois, O.; Pistolesi, F.; Collin, E. Scaling Laws for the Bifurcation Escape Rate in a Nanomechanical Resonator. {\it Phys. Rev. E} {\bf 2015}, 92, 050903(R).

\bibitem{allan} Allan, D. W. Time And Frequency (Time-Domain) Characterization, Estimation, and Prediction of Precision Clocks and Oscillators. {\it IEEE Trans. Ultrason. Eng.} {\bf 1987}, 34, 647-654.

\bibitem{martialthesis} Defoort, M. In {\it Non-linear dynamics in nano-electromechanical systems at low temperatures}; PhD Thesis: Universit\'e Grenoble Alpes, 2014.

\bibitem{nowick} Nowick, A.; Berry, B. In {\it Anelastic Relaxation in Crystalline Solids}; Academic Press: New York, 1972.

}
\end{thebibliography}

\begin{figure}[h!]		 
			 \includegraphics[width=8cm]{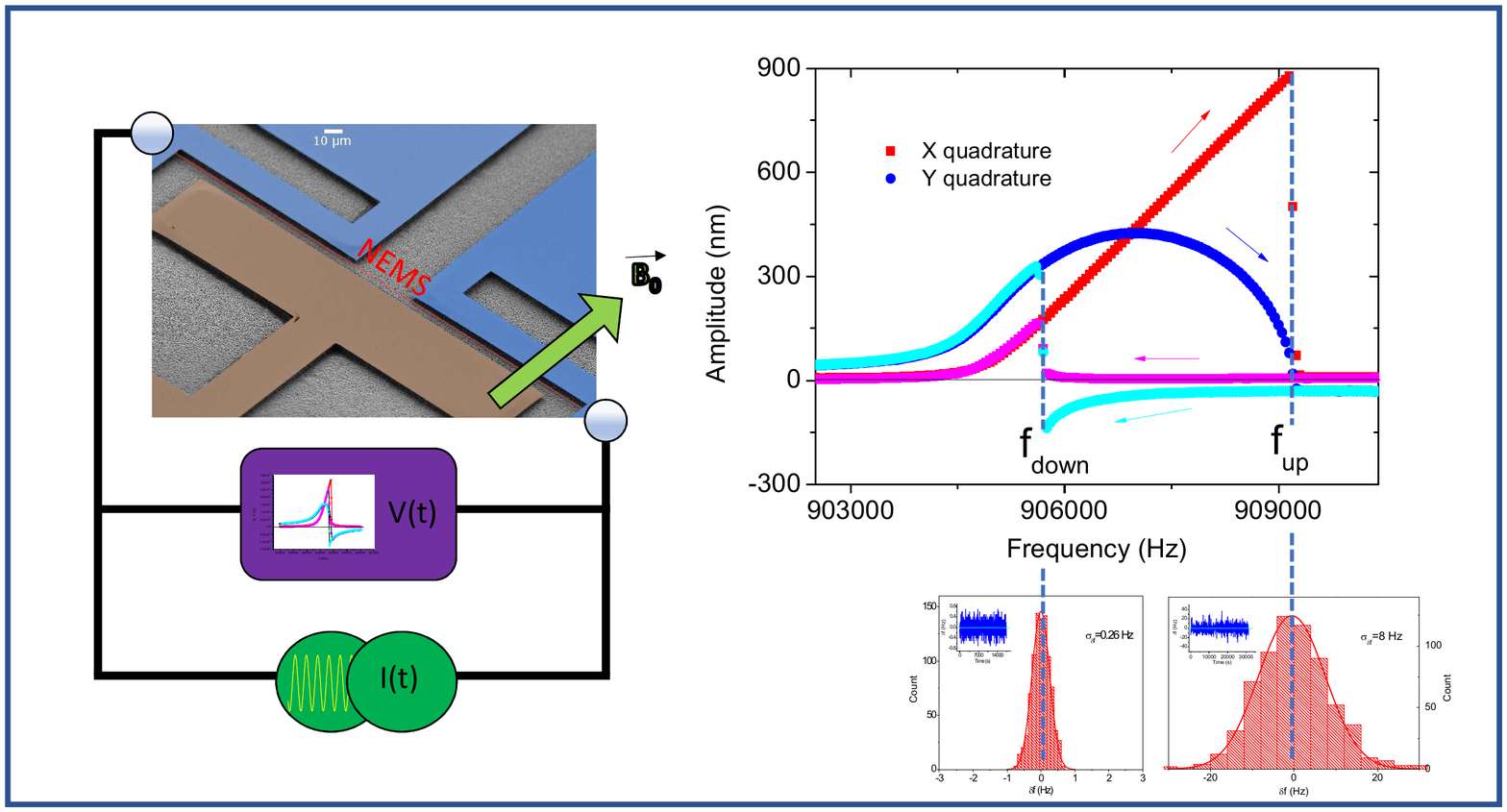}
\caption{\small{For Table Of Contents Only.}}
\label{fig:TOC}
\end{figure}	


%
%
%
%
%
%
%
%
%
%
%


\end{document}